\documentclass[journal]{IEEEtran}
\usepackage{amsmath}
\usepackage{amsfonts}
\usepackage{amssymb}
\usepackage{amstext}
\usepackage{amsthm}
\usepackage{gensymb}
\usepackage{graphicx}
\usepackage{algorithm}
\usepackage{algorithmic}
\usepackage{color}
\usepackage{url}
\usepackage{hyperref}

\newtheorem{remark}{Remark}
\newcounter{RomanNumber}

\begin{document}

\title{Stochastic Maximum Likelihood Direction Finding in the Presence of Nonuniform Noise Fields}

\author{Ming-yan Gong and Bin Lyu\IEEEmembership{}

\thanks{}
\thanks{}
\thanks{}
}

\markboth{} {}

\maketitle

\begin{abstract}
In this letter, we employ and design the expectation--conditional maximization either (ECME) algorithm, a generalisation of the EM algorithm, for solving the maximum likelihood direction finding problem of stochastic sources, which may be correlated, in unknown nonuniform noise. Unlike alternating maximization, the ECME algorithm updates both the source and noise covariance matrix estimates by explicit formulas and can guarantee that both estimates are positive semi-definite and definite, respectively. Thus, the ECME algorithm is computationally efficient and operationally stable. Simulation results confirm the effectiveness of the algorithm.
\end{abstract}


\IEEEpeerreviewmaketitle

\section{Introduction}

\IEEEPARstart{I}{t} is well known that two source signal models are widely used in Cramer-Rao lower bound (CRLB) and maximum likelihood (ML) direction finding, i.e., the deterministic signal model, where the signals are deterministic and unknown, and the stochastic signal model, where the signals are Gaussian. For example, various CRLBs using both models have been derived \cite{bibitem1}--\cite{bibitem7}. But, the ML direction finding generally involves high-dimensional search algorithms for both models, which causes a significant increase in the computational complexity.

In order to reduce the computational complexity, two classic methods have been developed: alternating maximization (AM) \cite{bibitem8} and expectation--maximization (EM) \cite{bibitem9}--\cite{bibitem12} type algorithms. Early, these two methods are applied under uniform Gaussian noise, which decreases the number of parameters and simplifies the problem. However, the uniform noise model is unrealistic in many situations and numerous papers have considered nonuniform noise \cite{bibitem4}, \cite{bibitem13}--\cite{bibitem19}. In nonuniform noise, the covariance matrix still keeps a diagonal structure but the diagonal elements are no longer identical, which makes direction of arrival (DOA) estimation difficult. To tackle the problem of direction finding in unknown nonuniform noise, diverse subspace separation approaches based on the subspace technique have been proposed in the literature \cite{bibitem13}--\cite{bibitem18}.

For obtaining ML based solutions, AM and EM type algorithms have also been applied to this problem. However, the AM type algorithms usually require high-dimensional numerical search due to the noise nonuniformity at each iteration \cite{bibitem4}, \cite{bibitem19}, which leads to a heavy computational burden. Moreover, when considering Gaussian source signals, the AM algorithm presented in \cite{bibitem19} has one severe shortcoming: the source and noise covariance matrix estimates cannot be guaranteed to be positive semi-definite and definite, respectively. To this end, we have designed several computationally efficient EM type algorithms in \cite{bibitem20}, which only need low-dimensional (one or two-dimensional) numerical search at every iteration. In these EM type algorithms using the stochastic signal model, however, the sources must be uncorrelated. This restricts the use of stochastic ML direction finding in some situations, e.g., multipath conditions. As a consequence, efficient algorithms are in urgent needs to address this issue.

In this letter, we employ and design the expectation--conditional maximization either (ECME) algorithm \cite{bibitem21}, a generalisation of the EM algorithm, for solving the ML direction finding problem of stochastic sources, which may be correlated, in unknown nonuniform noise. Unlike the AM algorithm in \cite{bibitem19}, the ECME algorithm updates both the source and noise covariance matrix estimates by explicit formulas and can guarantee that both estimates are positive semi-definite and definite, respectively. Thus, the ECME algorithm is computationally efficient and operationally stable. Simulation results confirm the effectiveness of the algorithm.

\section{Problem Statement}

For simplicity, let a uniformly spaced linear array of $W$ sensors receive the plane wave(s) impinging from $V$ $(V<W)$ narrow-band source(s) of wavelength $\iota$. The distance between any adjacent sensors is $\iota/2$. We denote the direction associated with the $v$th source by $\beta_v\in(0,\pi)$ $(\mathrm{radian})$ and write the received signal as
\begin{equation}\label{eq1}     
\mathbf{r}(t)={\sum}_{v=1}^V\mathbf{a}(\beta_v)k_v(t)+\mathbf{j}(t)=\mathbf{A}(\boldsymbol{\beta})\mathbf{k}(t)+\mathbf{j}(t),
\end{equation}
where $\mathbf{a}(\beta_v)=[1~a_v~\cdots~a_v^{W-1}]^T$, $a_v=\exp\big(-\jmath\pi\cos(\beta_v)\big)$, $[\cdot]^T$ denotes transposition, $\jmath=\sqrt{-1}$, $k_v(t)$ is the signal with respect to the $v$th source, and $\mathbf{j}(t)$ means nonuniform complex Gaussian noise of zero mean and covariance $\mathbf{Q}$, i.e., $\mathbf{j}(t)\sim\mathcal{CN}(\mathbf{0},\mathbf{Q})$. Here, $\mathbf{Q}$ is diagonal and expressed as
\begin{equation}
\mathbf{Q}=\mathrm{diag}\{\boldsymbol{\delta}\},\boldsymbol{\delta}=[\delta_1~\cdots~\delta_W]^T>\mathbf{0},\nonumber
\end{equation}
where $\mathbf{Q}$ is positive definite, i.e., $\mathbf{Q}\succ\mathbf{0}_W$ ($\mathbf{0}_W$ is the $W\times W$ zero matrix).
Furthermore, if $\delta_1=\cdots=\delta_W=\delta>0$, $\mathbf{Q}=\delta\mathbf{I}_W$ ($\mathbf{I}_W$ is the $W\times W$ identity matrix), which makes the noise uniform. In \eqref{eq1}, $\mathbf{A}(\boldsymbol{\beta})=[\mathbf{a}(\beta_1)~\cdots~\mathbf{a}(\beta_V)]$ is the array manifold matrix, $\boldsymbol{\beta}=[\beta_1~\cdots~\beta_V]^T\in\boldsymbol{\Gamma}$ with $\boldsymbol{\Gamma}=(0,\pi)^V$, and $\mathbf{k}(t)=[k_1(t)~\cdots~k_V(t)]^T$. For notational convenience, we use $\mathbf{A}$ instead of $\mathbf{A}(\boldsymbol{\beta})$ hereafter.

We consider Gaussian source signals, which may be correlated, and have $\mathbf{k}(t)\sim\mathcal{CN}(\mathbf{0},\mathbf{O})$, where $\mathbf{O}$ is the source covariance matrix and positive semi-definite, i.e., $\mathbf{O}\succeq\mathbf{0}_V$.
Let the source(s) be uncorrelated with the noise such that
\begin{eqnarray}
\mathbf{r}(t)\sim\mathcal{CN}(\mathbf{0},\mathbf{G}),\mathbf{G}=\mathbf{A}\mathbf{O}\mathbf{A}^H+\mathbf{Q}\succ\mathbf{0}_W,\nonumber
\end{eqnarray}
where $[\cdot]^H$ is conjugate transposition. On this foundation, the log-likelihood function (LLF) of $L$ statistically independent snapshot(s) can be formulated as
\begin{eqnarray}\label{eq2}       
\mathcal{J}(\boldsymbol{\beta},\boldsymbol{\rho},\boldsymbol{\delta})=&
\sum_{t=1}^L\log\mathnormal{p}\big(\mathbf{r}(t);\boldsymbol{\beta},\boldsymbol{\rho},\boldsymbol{\delta}\big)\nonumber \\
=&f-L\big(\log\vert\mathbf{G}\vert+\mathrm{trace}\big[\mathbf{G}^{-1}\hat{\mathbf{R}}\big]\big),
\end{eqnarray}
where $\vert\cdot\vert$, $\mathrm{trace}[\cdot]$, and $(\cdot)^{-1}$ denote determinant, trace, and inversion, respectively. In \eqref{eq2}, $f$ is a constant, $\hat{\mathbf{R}}=(1/L)\sum_{t=1}^L\mathbf{r}(t)\mathbf{r}^H(t)$ means the covariance matrix of snapshots. Moreover,
\begin{eqnarray}
\boldsymbol{\rho}=&\big([\mathbf{O}]_{1,1},\dots,[\mathbf{O}]_{V,V},\mathrm{Re}\big\{[\mathbf{O}]_{1,2}\big\},\mathrm{Im}\big\{[\mathbf{O}]_{1,2}\big\},\dots,\nonumber\\
&\mathrm{Re}\big\{[\mathbf{O}]_{V-1,V}\big\},\mathrm{Im}\big\{[\mathbf{O}]_{V-1,V}\big\}\big),\nonumber
\end{eqnarray}
where $[\mathbf{O}]_{p,q}$ is the $(p,q)$th element of $\mathbf{O}$, $\mathrm{Re}\{a\}$ and $\mathrm{Im}\{a\}$ represent the real part and imaginary part of $a$, respectively. Consequently, the ML based DOA estimation problem is
\begin{eqnarray}\label{eq3}       
\max_{\boldsymbol{\beta}\in\boldsymbol{\Gamma},\mathbf{O}\succeq\mathbf{0}_V,\boldsymbol{\delta}>\mathbf{0}}
\mathcal{J}(\boldsymbol{\beta},\boldsymbol{\rho},\boldsymbol{\delta}).
\end{eqnarray}

We assume $\mathcal{R}[\mathbf{A}]=V$, where $\mathcal{R}[\mathbf{A}]$ is the rank of $\mathbf{A}$, and can thus eliminate $\mathbf{O}$ in \eqref{eq3} by \cite{bibitem22}
\begin{eqnarray}\label{eq4}       
\hat{\mathbf{O}}(\boldsymbol{\beta},\boldsymbol{\delta})=\big(\tilde{\mathbf{A}}^H\tilde{\mathbf{A}}\big)^{-1}\tilde{\mathbf{A}}^H\big(\widetilde{\mathbf{R}}-\mathbf{I}_W\big)
\tilde{\mathbf{A}}\big(\tilde{\mathbf{A}}^H\tilde{\mathbf{A}}\big)^{-1},
\end{eqnarray}
where $\mathbf{Q}^{-1/2}=\mathrm{diag}\{1/\sqrt{\delta_1},\dots,1/\sqrt{\delta_W}\}$,
\begin{eqnarray}
\tilde{\mathbf{A}}=\mathbf{Q}^{-1/2}\mathbf{A},~\text{and}~
\widetilde{\mathbf{R}}=\mathbf{Q}^{-1/2}\hat{\mathbf{R}}\mathbf{Q}^{-1/2}.\nonumber
\end{eqnarray}
In other words, $\boldsymbol{\rho}$ can be estimated using the estimates of $\boldsymbol{\beta}$ and $\boldsymbol{\delta}$. Based on \eqref{eq4}, $\mathbf{G}$ is rewritten as
\begin{eqnarray}
\mathbf{G}=\mathbf{Q}^{1/2}\big(\boldsymbol{\Pi}_{\tilde{\mathbf{A}}}\widetilde{\mathbf{R}}\boldsymbol{\Pi}_{\tilde{\mathbf{A}}}
+\boldsymbol{\Pi}^{\perp}_{\tilde{\mathbf{A}}}\big)\mathbf{Q}^{1/2},\nonumber
\end{eqnarray}
where $\boldsymbol{\Pi}_{\tilde{\mathbf{A}}}=\tilde{\mathbf{A}}\big(\tilde{\mathbf{A}}^H\tilde{\mathbf{A}}\big)^{-1}\tilde{\mathbf{A}}^H$
and $\boldsymbol{\Pi}^{\perp}_{\tilde{\mathbf{A}}}=\mathbf{I}_W-\boldsymbol{\Pi}_{\tilde{\mathbf{A}}}$. Then, problem \eqref{eq3} is reduced to \cite{bibitem22}
\begin{eqnarray}\label{eq5}       
\min_{\boldsymbol{\beta}\in\boldsymbol{\Gamma},\boldsymbol{\delta}>\mathbf{0}}\mathcal{H}(\boldsymbol{\beta},\boldsymbol{\delta})=
\log\big\vert\mathbf{Q}^{1/2}\big(\boldsymbol{\Pi}_{\tilde{\mathbf{A}}}\widetilde{\mathbf{R}}\boldsymbol{\Pi}_{\tilde{\mathbf{A}}}
+\boldsymbol{\Pi}^{\perp}_{\tilde{\mathbf{A}}}\big)\mathbf{Q}^{1/2}\big\vert\nonumber\\
+\mathrm{trace}\big[\big(\boldsymbol{\Pi}_{\tilde{\mathbf{A}}}\widetilde{\mathbf{R}}\boldsymbol{\Pi}_{\tilde{\mathbf{A}}}
+\boldsymbol{\Pi}^{\perp}_{\tilde{\mathbf{A}}}\big)^{-1}\widetilde{\mathbf{R}}\big].
\end{eqnarray}
In particular, if the noise is uniform Gaussian noise, problem \eqref{eq5} can be further reduced to \cite{bibitem23}
\begin{eqnarray}\label{eq6}       
\min_{\boldsymbol{\beta}\in\boldsymbol{\Gamma}}\mathcal{G}(\boldsymbol{\beta})=
\big\vert\mathbf{A}\hat{\mathbf{O}}(\boldsymbol{\beta})\mathbf{A}^H+\hat{\delta}(\boldsymbol{\beta})\mathbf{I}_W\big\vert,
\end{eqnarray}
where
\begin{eqnarray}
\hat{\delta}(\boldsymbol{\beta})&=&\mathrm{trace}\big[\big(\mathbf{I}_W-\mathbf{A}(\mathbf{A}^H\mathbf{A})^{-1}\mathbf{A}^H\big)\hat{\mathbf{R}}\big]/(W-V),\nonumber\\
\hat{\mathbf{O}}(\boldsymbol{\beta})&=&(\mathbf{A}^H\mathbf{A})^{-1}\mathbf{A}^H\big(\hat{\mathbf{R}}-\hat{\delta}(\boldsymbol{\beta})\mathbf{I}_W\big)
\mathbf{A}(\mathbf{A}^H\mathbf{A})^{-1}.\nonumber
\end{eqnarray}
Unfortunately, it is very difficult to reduce problem \eqref{eq5} to some problems with fewer parameters under nonuniform Gaussian noise. Of course, applying gradient type algorithms to search the solution of problem \eqref{eq5} is computationally intensive due to the search space of dimension $W+V$ and the complexity of $\mathcal{H}(\boldsymbol{\beta},\boldsymbol{\delta})$.

In fact, when a direct maximization over all parameters is intractable, AM can always be utilized. As stated before, the authors in \cite{bibitem19} have presented an AM algorithm consisting of two steps at every iteration for problem \eqref{eq3}. Specifically, the first step obtains $\boldsymbol{\delta}^{(d)}$, the estimate of $\boldsymbol{\delta}$ at the $d$th iteration, by a gradient based algorithm, which is called the ``modified inverse iteration algorithm'' and satisfies
\begin{eqnarray}\label{eq7}  
\mathcal{J}(\boldsymbol{\beta}^{(d-1)},\boldsymbol{\rho}^{(d-1)},\boldsymbol{\delta}^{(d)})
\ge\mathcal{J}(\boldsymbol{\beta}^{(d-1)},\boldsymbol{\rho}^{(d-1)},\boldsymbol{\delta}^{(d-1)}),
\end{eqnarray}
where $[\cdot]^{(0)}$ means an initial estimate.
Then, the second step simultaneously obtains $\boldsymbol{\beta}^{(d)}$ and $\boldsymbol{\rho}^{(d)}$ by
\begin{eqnarray}\label{eq8}       
(\boldsymbol{\beta}^{(d)},\boldsymbol{\rho}^{(d)})=\arg\max_{\boldsymbol{\beta}\in\boldsymbol{\Gamma},\mathbf{O}\succeq\mathbf{0}_V}
\mathcal{J}(\boldsymbol{\beta},\boldsymbol{\rho},\boldsymbol{\delta}^{(d)}),
\end{eqnarray}
which is solved in a separable manner, i.e.,
\begin{eqnarray}\label{eq9}       
\boldsymbol{\beta}^{(d)}&=&\arg\min_{\boldsymbol{\beta}\in\boldsymbol{\Gamma}}\mathcal{H}(\boldsymbol{\beta},\boldsymbol{\delta}^{(d)}),\\
\mathbf{O}^{(d)}&=&\hat{\mathbf{O}}(\boldsymbol{\beta}^{(d)},\boldsymbol{\delta}^{(d)}).\label{eq10}
\end{eqnarray}

However, the AM algorithm has two drawbacks: 1) obtaining $\boldsymbol{\delta}^{(d)}$ and $\boldsymbol{\beta}^{(d)}$ is computationally expensive, 2) $\mathbf{Q}^{(d)}\succ\mathbf{0}_W$ (or $\boldsymbol{\delta}^{(d)}>\mathbf{0}$) and $\mathbf{O}^{(d)}\succeq\mathbf{0}_V$ cannot be guaranteed \cite{bibitem24}, \cite{bibitem25}. For efficiently obtaining the ML estimate of $\boldsymbol{\beta}$ in \eqref{eq3}, we employ and design the ECME algorithm in the next section.

\section{ECME Algorithm}

Existing EM type algorithms for stochastic ML direction finding are only applicable to uncorrelated sources \cite{bibitem9}, \cite{bibitem12}, \cite{bibitem20}, i.e., $\mathbf{O}$ is diagonal. In this section, we employ the ECME algorithm \cite{bibitem21}, a generalisation of the EM algorithm, to solve problem \eqref{eq3} associated with correlated sources.

\subsection{Procedure}

The source(s) in \eqref{eq1} may be correlated, so we choose $\mathbf{K}=[\mathbf{k}(1)~\cdots~\mathbf{k}(L)]$ and $\mathbf{J}=[\mathbf{j}(1)~\cdots~\mathbf{j}(L)]$ as augmented data. We express the augmented-data LLF as
\begin{eqnarray}\label{eq11}       
\mathcal{M}(\mathbf{K},\mathbf{J},\boldsymbol{\rho},\boldsymbol{\delta})=&\sum_{t=1}^L\big[\log\mathnormal{p}(\mathbf{k}(t);\boldsymbol{\rho})+
\log\mathnormal{p}(\mathbf{j}(t);\boldsymbol{\delta})\big]\nonumber \\
=&h-L\big(\log\vert\mathbf{O}\vert+\mathrm{trace}\big[\mathbf{O}^{-1}\hat{\mathbf{N}}_k\big]\big)+\nonumber \\
&f-L\big(\log\vert\mathbf{Q}\vert+\mathrm{trace}\big[\mathbf{Q}^{-1}\hat{\mathbf{N}}_j\big]\big),
\end{eqnarray}
where $h$ is a constant, $\hat{\mathbf{N}}_k=(1/L)\sum_{t=1}^L\mathbf{k}(t)\mathbf{k}^H(t)$, and $\hat{\mathbf{N}}_j=(1/L)\sum_{t=1}^L\mathbf{j}(t)\mathbf{j}^H(t)$. With \eqref{eq11}, we first construct the EM algorithm, whose expectation step and maximization step at the $d$th iteration are derived below. Let $\mathcal{E}\{\cdot\}$ and $\mathcal{D}\{\cdot\}$ represent expectation and covariance, respectively.

\subsubsection{Expectation Step}

Compute the conditional expectation of the augmented-data LLF, i.e.,
\begin{eqnarray}\label{eq12}       
&\mathcal{M}\big(\boldsymbol{\rho},\boldsymbol{\delta};\boldsymbol{\Omega}^{(d-1)}\big)
=\mathcal{E}\big\{\mathcal{M}(\mathbf{K},\mathbf{J},\boldsymbol{\rho},\boldsymbol{\delta})\big|
\mathbf{F};\boldsymbol{\Omega}^{(d-1)}\big\}\nonumber\\
=&h-L\big(\log\vert\mathbf{O}\vert+\mathrm{trace}\big[\mathbf{O}^{-1}\hat{\mathbf{N}}^{(d)}_k\big]\big)+\nonumber\\
&f-L\big(\log\vert\mathbf{Q}\vert+\mathrm{trace}\big[\mathbf{Q}^{-1}\hat{\mathbf{N}}^{(d)}_j\big]\big)
\end{eqnarray}
with $\boldsymbol{\Omega}=(\boldsymbol{\beta},\boldsymbol{\rho},\boldsymbol{\delta})$ and $\mathbf{F}=[\mathbf{r}(1)~\cdots~\mathbf{r}(L)]$. Moreover,
\begin{eqnarray}\label{eq13}       
\hat{\mathbf{N}}^{(d)}_k=&\mathcal{E}\big\{\hat{\mathbf{N}}_k\big\vert\mathbf{F};\boldsymbol{\Omega}^{(d-1)}\big\}
=[\mathbf{H}^{(d-1)}]^H\hat{\mathbf{R}}\mathbf{H}^{(d-1)}+\nonumber\\
&\mathbf{O}^{(d-1)}-[\mathbf{H}^{(d-1)}]^H\mathbf{G}^{(d-1)}\mathbf{H}^{(d-1)}\succeq\mathbf{0}_V,\\
\hat{\mathbf{N}}^{(d)}_j=&\mathcal{E}\big\{\hat{\mathbf{N}}_j\big\vert\mathbf{F};\boldsymbol{\Omega}^{(d-1)}\big\}\label{eq14}\nonumber\\
=&\mathbf{Q}^{(d-1)}[\mathbf{G}^{(d-1)}]^{-1}\hat{\mathbf{R}}[\mathbf{G}^{(d-1)}]^{-1}\mathbf{Q}^{(d-1)}+\nonumber\\
&\mathbf{Q}^{(d-1)}-\mathbf{Q}^{(d-1)}[\mathbf{G}^{(d-1)}]^{-1}\mathbf{Q}^{(d-1)}\succeq\mathbf{0}_W,
\end{eqnarray}
where $\mathbf{H}^{(d-1)}=[\mathbf{G}^{(d-1)}]^{-1}\mathbf{A}^{(d-1)}\mathbf{O}^{(d-1)}$,
the conditional distributions of $\mathbf{k}(t)$ and $\mathbf{j}(t)$ can be obtained in \cite{bibitem26}, and
\begin{eqnarray}
\mathcal{E}\big\{\mathbf{k}(t)\big\vert\mathbf{F};\boldsymbol{\Omega}^{(d-1)}\big\}
=&[\mathbf{H}^{(d-1)}]^H\mathbf{r}(t),\nonumber\\
\mathcal{D}\big\{\mathbf{k}(t)\big\vert\mathbf{F};\boldsymbol{\Omega}^{(d-1)}\big\}
=&\mathbf{O}^{(d-1)}-\nonumber\\
&[\mathbf{H}^{(d-1)}]^H\mathbf{G}^{(d-1)}\mathbf{H}^{(d-1)}\succeq\mathbf{0}_V,\nonumber\\
\mathcal{E}\big\{\mathbf{j}(t)\big\vert\mathbf{F};\boldsymbol{\Omega}^{(d-1)}\big\}
=&\mathbf{Q}^{(d-1)}[\mathbf{G}^{(d-1)}]^{-1}\mathbf{r}(t),\nonumber\\
\mathcal{D}\big\{\mathbf{j}(t)\big\vert\mathbf{F};\boldsymbol{\Omega}^{(d-1)}\big\}
=&\mathbf{Q}^{(d-1)}-\nonumber\\
&\mathbf{Q}^{(d-1)}[\mathbf{G}^{(d-1)}]^{-1}\mathbf{Q}^{(d-1)}\succeq\mathbf{0}_W.\nonumber
\end{eqnarray}

\subsubsection{Maximization Step}

Obtain $\boldsymbol{\rho}^{(d)}$ and $\boldsymbol{\delta}^{(d)}$ by maximizing $\mathcal{M}(\boldsymbol{\rho},\boldsymbol{\delta};\boldsymbol{\Omega}^{(d-1)})$ with respect to $\boldsymbol{\rho}$ and $\boldsymbol{\delta}$, which leads to the two parallel subproblems
\begin{eqnarray}\label{eq15}       
\min_{\mathbf{O}\succeq\mathbf{0}_V}\log\vert\mathbf{O}\vert+\mathrm{trace}\big[\mathbf{O}^{-1}\hat{\mathbf{N}}^{(d)}_k\big],\\
\min_{\mathbf{Q}\succ\mathbf{0}_W}\log\vert\mathbf{Q}\vert+\mathrm{trace}\big[\mathbf{Q}^{-1}\hat{\mathbf{N}}^{(d)}_j\big].
\end{eqnarray}
$\boldsymbol{\rho}^{(d)}$ and $\boldsymbol{\delta}^{(d)}$ are simultaneously obtained by
\begin{eqnarray}\label{eq17}       
\mathbf{O}^{(d)}&=&\hat{\mathbf{N}}^{(d)}_k\succeq\mathbf{0}_V,\\
\delta_w^{(d)}&=&\left\{\label{eq18}
\begin{array}{ll}
[\hat{\mathbf{N}}^{(d)}_j]_{w,w},&[\hat{\mathbf{N}}^{(d)}_j]_{w,w}>0,\\
\delta_w^{(d-1)}/2,&[\hat{\mathbf{N}}^{(d)}_j]_{w,w}=0,
\end{array}
\right.
\forall w.
\end{eqnarray}

From \eqref{eq17} and \eqref{eq18}, we have the monotonicity of generalized EM algorithms \cite{bibitem27}, i.e.,
\begin{eqnarray}\label{eq19}       
\mathcal{J}\big(\boldsymbol{\beta}^{(d-1)},\boldsymbol{\rho}^{(d)},\boldsymbol{\delta}^{(d)}\big)
\ge\mathcal{J}\big(\boldsymbol{\beta}^{(d-1)},\boldsymbol{\rho}^{(d-1)},\boldsymbol{\delta}^{(d-1)}\big).
\end{eqnarray}
Obviously, $\boldsymbol{\beta}^{(d)}$ is not obtained at the $d$th iteration of the EM algorithm.

\subsubsection{Conditional Maximization Step}

In order to obtain $\boldsymbol{\beta}^{(d)}$, we now add a conditional maximization step at this iteration. Considering the monotonicity
\begin{eqnarray}\label{eq20} 
\mathcal{J}\big(\boldsymbol{\beta}^{(d)},\boldsymbol{\rho}^{(d)},\boldsymbol{\delta}^{(d)}\big)
\ge\mathcal{J}\big(\boldsymbol{\beta}^{(d-1)},\boldsymbol{\rho}^{(d)},\boldsymbol{\delta}^{(d)}\big),\boldsymbol{\beta}^{(d)}\in\boldsymbol{\Gamma},
\end{eqnarray}
we can design this step as
\begin{eqnarray}\label{eq21}       
\boldsymbol{\beta}^{(d)}=\arg\max_{\boldsymbol{\beta}\in\boldsymbol{\Gamma}}
\mathcal{J}\big(\boldsymbol{\beta},\boldsymbol{\rho}^{(d)},\boldsymbol{\delta}^{(d)}\big),
\end{eqnarray}
or use a gradient type algorithm to obtain $\boldsymbol{\beta}^{(d)}$ based on \eqref{eq20}, e.g., \textbf{Algorithm 1} in the next section.
Due to the additional step unrelated to augmented data, the above EM algorithm becomes the ECME algorithm \cite{bibitem21}.

\subsection{Stability and Complexity}

The stable operation of the algorithm requires $\mathbf{Q}^{(d)}\succ\mathbf{0}_W$ (or $\boldsymbol{\delta}^{(d)}>\mathbf{0}$) and $\mathbf{O}^{(d)}\succeq\mathbf{0}_V$ for $d\ge0$, so we give the following remark.

\begin{remark}\label{re1}
In the ECME algorithm, $\mathbf{Q}^{(d)}\succ\mathbf{0}_W$ (or $\boldsymbol{\delta}^{(d)}>\mathbf{0}$) and $\mathbf{O}^{(d)}\succeq\mathbf{0}_V$ for $d\ge1$ if $\mathbf{Q}^{(0)}\succ\mathbf{0}_W$ and $\mathbf{O}^{(0)}\succeq\mathbf{0}_V$.
\end{remark}

\begin{proof}
We utilize the mathematical induction method. If $\mathbf{Q}^{(u)}\succ\mathbf{0}_W$ (or $\boldsymbol{\delta}^{(u)}>\mathbf{0}$) and $\mathbf{O}^{(u)}\succeq\mathbf{0}_V$, we have $\mathbf{G}^{(u)}=\mathbf{A}^{(u)}\mathbf{O}^{(u)}[\mathbf{A}^{(u)}]^H+\mathbf{Q}^{(u)}\succ\mathbf{0}_W$, which leads to $\hat{\mathbf{N}}^{(u+1)}_j\succeq\mathbf{0}_W$ in (14) and then in (18) $\delta_w^{(u+1)}>0,\forall w$, i.e., $\mathbf{Q}^{(u+1)}\succ\mathbf{0}_W$. Furthermore, $\mathbf{O}^{(u+1)}=\hat{\mathbf{N}}^{(u+1)}_k\succeq\mathbf{0}_V$ in (13). The proof is completed.
\end{proof}

\textbf{Remark \ref{re1}} indicates that when $\mathbf{Q}^{(0)}\succ\mathbf{0}_W$ and $\mathbf{O}^{(0)}\succeq\mathbf{0}_V$ in the ECME algorithm, $\boldsymbol{\rho}^{(d)}$ and $\boldsymbol{\delta}^{(d)}$ obtained at the $d$th iteration are in the parameter spaces, respectively. Hence, the ECME algorithm is operationally stable.

Since $\boldsymbol{\rho}^{(d)}$ and $\boldsymbol{\delta}^{(d)}$ are obtained by the explicit formulas in \eqref{eq17} and \eqref{eq18}, the computational complexity of the ECME algorithm is dominated by obtaining $\boldsymbol{\beta}^{(d)}$ in \eqref{eq20}. Compared with the AM algorithm in \cite{bibitem19}, the ECME algorithm is thus computationally efficient.

\subsection{Limit Point}

According to \cite{bibitem21}, \cite{bibitem28}, we know that the ECME algorithm satisfies certain regularity conditions and always converges to a stationary point of $\mathcal{J}(\boldsymbol{\beta},\boldsymbol{\rho},\boldsymbol{\delta})$. Unfortunately, $\mathcal{J}(\boldsymbol{\beta},\boldsymbol{\rho},\boldsymbol{\delta})$ tends to have multiple stationary points and the limit point of the ECME algorithm may be an undesirable stationary point.
To deal with this issue, we need to provide an accurate initial point. Following the method in \cite{bibitem19}, we can assume that the noise is uniform and then evaluate $\mathcal{G}(\boldsymbol{\beta})$ in (6) on a coarse $V$-dimensional grid to find a grid point, close to the global minimum of $\mathcal{G}(\boldsymbol{\beta})$, as $\boldsymbol{\beta}^{(0)}$ of the ECME algorithm. Besides, we can use the estimate of $\boldsymbol{\beta}$, obtained by a subspace \cite{bibitem29} or sparse representation \cite{bibitem30} based algorithm, as $\boldsymbol{\beta}^{(0)}$ due to the higher accuracy of the stochastic ML estimate of $\boldsymbol{\beta}$ \cite{bibitem2}.

On the boundary of the positive semi-definite region of $\boldsymbol{\rho}$, i.e., the set $\eth=\{\boldsymbol{\rho}\mid\mathbf{O}\succeq\mathbf{0}_V,\mathcal{R}[\mathbf{O}]<V\}$, we give the following remark. Let $\mathcal{N}[\mathbf{O}]$ denote the null space of $\mathbf{O}$.

\begin{remark} \label{re2}
In the ECME algorithm, $\mathcal{N}[\mathbf{O}^{(d)}]=\mathcal{N}[\mathbf{O}^{(0)}]$ for $d\ge1$ if $\mathbf{Q}^{(0)}\succ\mathbf{0}_W$ and $\mathbf{O}^{(0)}\succeq\mathbf{0}_V$.
\end{remark}

\begin{proof}
From \textbf{Remark \ref{re1}}, we first know that $\mathbf{Q}^{(d)}\succ\mathbf{0}_W$, $\mathbf{O}^{(d)}\succeq\mathbf{0}_V$, and $\mathbf{G}^{(d)}\succ\mathbf{0}_W$ for $d\ge0$ due to $\mathbf{Q}^{(0)}\succ\mathbf{0}_W$ and $\mathbf{O}^{(0)}\succeq\mathbf{0}_V$. Then, a proof by the mathematical induction method is given.

If $\mathbf{O}^{(u)}\mathbf{v}=\mathbf{0}$, we have $\mathbf{O}^{(u+1)}\mathbf{v}=\hat{\mathbf{N}}_k^{(u+1)}\mathbf{v}=\mathbf{0}$ in \eqref{eq13} and thus $\mathcal{N}[\mathbf{O}^{(u)}]\subseteq\mathcal{N}[\mathbf{O}^{(u+1)}]$. Furthermore, if $\mathbf{O}^{(u+1)}\mathbf{v}=\hat{\mathbf{N}}_k^{(u+1)}\mathbf{v}=\mathbf{0}$, we have $\mathbf{v}^H\hat{\mathbf{N}}_k^{(u+1)}\mathbf{v}=0$ and in \eqref{eq13}
\begin{eqnarray}\label{eq22} 
&\mathbf{v}^H[\mathbf{H}^{(u)}]^H\hat{\mathbf{R}}\mathbf{H}^{(u)}\mathbf{v}=0\Rightarrow[\mathbf{H}^{(u)}]^H\hat{\mathbf{R}}\mathbf{H}^{(u)}\mathbf{v}=\mathbf{0},\\
\label{eq23}&\mathbf{v}^H\big(\mathbf{O}^{(u)}-[\mathbf{H}^{(u)}]^H\mathbf{G}^{(u)}\mathbf{H}^{(u)}\big)\mathbf{v}=0\nonumber\\
&\Rightarrow\big(\mathbf{O}^{(u)}-[\mathbf{H}^{(u)}]^H\mathbf{G}^{(u)}\mathbf{H}^{(u)}\big)\mathbf{v}=\mathbf{0}.
\end{eqnarray}
To proceed, we use the matrix inversion formula \cite{bibitem22}
\begin{eqnarray}\label{eq24} 
\mathbf{G}^{-1}=\mathbf{Q}^{-1/2}\big[\mathbf{I}_W-\tilde{\mathbf{A}}(\mathbf{O}\tilde{\mathbf{A}}^H\tilde{\mathbf{A}}
+\mathbf{I}_V)^{-1}\mathbf{O}\tilde{\mathbf{A}}^H\big]\mathbf{Q}^{-1/2}
\end{eqnarray}
and obtain
\begin{eqnarray}\label{eq25} 
\mathbf{O}-\mathbf{H}^H\mathbf{G}\mathbf{H}=
(\mathbf{O}\tilde{\mathbf{A}}^H\tilde{\mathbf{A}}+\mathbf{I}_V)^{-1}\mathbf{O},
\end{eqnarray}
which suggests
$\mathcal{N}[\mathbf{O}]=\mathcal{N}[\mathbf{O}-\mathbf{H}^H\mathbf{G}\mathbf{H}]$ and $\mathcal{N}[\mathbf{O}^{(u)}]=\mathcal{N}[\mathbf{O}^{(u)}-[\mathbf{H}^{(u)}]^H\mathbf{G}^{(u)}\mathbf{H}^{(u)}]$. Accordingly, $\mathbf{O}^{(u)}\mathbf{v}=\mathbf{0}$ in \eqref{eq22} and \eqref{eq23}, leading to $\mathcal{N}[\mathbf{O}^{(u)}]\supseteq\mathcal{N}[\mathbf{O}^{(u+1)}]$. Finally, combining $\mathcal{N}[\mathbf{O}^{(u)}]\subseteq\mathcal{N}[\mathbf{O}^{(u+1)}]$ and $\mathcal{N}[\mathbf{O}^{(u)}]\supseteq\mathcal{N}[\mathbf{O}^{(u+1)}]$ obtains $\mathcal{N}[\mathbf{O}^{(u)}]=\mathcal{N}[\mathbf{O}^{(u+1)}]$. The proof is completed.
\end{proof}

\textbf{Remark \ref{re2}} indicates that if $\boldsymbol{\rho}^{(0)}$ in the ECME algorithm is on the boundary, i.e., $\boldsymbol{\rho}^{(0)}\in\eth$ and $\mathcal{N}[\mathbf{O}^{(0)}]$ is nonempty, the limit point of $\boldsymbol{\rho}$ is also on the boundary. Hence, let $(\boldsymbol{\beta}^{\ast},\boldsymbol{\rho}^{\ast},\boldsymbol{\delta}^{\ast})$ denote the solution of problem \eqref{eq3} and if $\boldsymbol{\rho}^{\ast}\in\eth$, we may need to estimate $\mathcal{N}[\mathbf{O}^{\ast}]$ before implementing the ECME algorithm. Fortunately, $(\boldsymbol{\beta}^{\ast},\boldsymbol{\rho}^{\ast},\boldsymbol{\delta}^{\ast})$ is always an interior point of the parameter space (i.e., $\boldsymbol{\rho}^{\ast}\notin\eth$, $\mathbf{O}^{\ast}\succ\mathbf{0}_V$, and $\mathcal{N}[\mathbf{O}^{\ast}]$ is empty) in practice even if the true value of $\boldsymbol{\rho}$ is on the boundary. As a result, we can always adopt $\mathbf{O}^{(0)}\succ\mathbf{0}_V$ in the ECME algorithm, e.g., the simulation results in Fig. \ref{fig1} related to coherent sources.

\section{Simulation Results}

\begin{figure}[t] \centering
\includegraphics[scale=0.55]{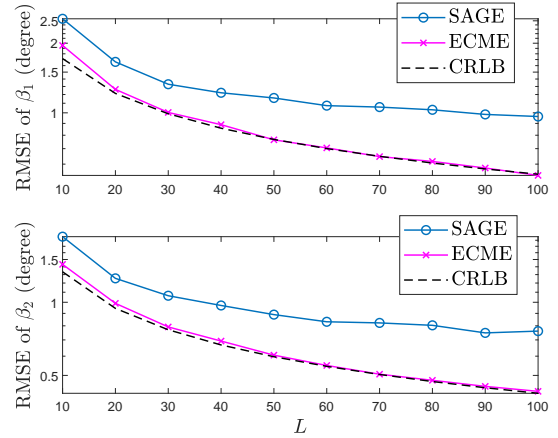}
\vspace{0cm}\caption{Relationship between the RMSE performance of the ECME algorithm and the CRLB. $\beta^{(0)}_1=45\degree$, $\textbf{O}^{(0)}=\mathbf{I}_V$, $\beta^{(0)}_2=95\degree$, and $\mathbf{Q}^{(0)}=\mathbf{I}_W$. \label{fig1}}\vspace{0cm}
\end{figure}

\begin{figure}[t] \centering
\includegraphics[scale=0.55]{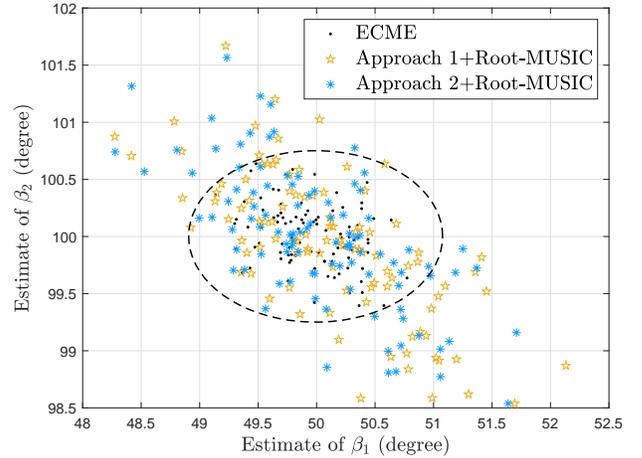}
\vspace{0cm}\caption{Estimates of $(\beta_1,\beta_2)$ obtained from the ECME and two subspace based algorithms under $100$ independent trials. $\beta^{(0)}_1=45\degree$, $\beta^{(0)}_2=95\degree$, $\mathbf{O}^{(0)}=\mathbf{I}_V$, $L=100$, and $\mathbf{Q}^{(0)}=\mathbf{I}_W$. \label{fig2}}\vspace{0cm}
\end{figure}

Simulation results are provided to confirm the effectiveness of the ECME algorithm, i.e., the ECME algorithm is able to obtain the ML estimate of $\boldsymbol{\beta}$ in \eqref{eq3}. We set $V=2$, $\beta_1=50\degree$, $W=6$, $\beta_2=100\degree$, and $\boldsymbol{\delta}=[1~2~3~4~2~10]^T$. \textbf{Algorithm \ref{alg1}} is used to obtain $\boldsymbol{\beta}^{(d)}$ in (20) and $\Vert\boldsymbol{\beta}^{(u+1)}-\boldsymbol{\beta}^{(u)}\Vert_2\le0.001\degree$ is adopted as the stopping criterion. The ECME algorithm is given an accurate initial point for obtaining the ML estimate of $\boldsymbol{\beta}$. In Figs. \ref{fig1} and \ref{fig2}, we, respectively, consider the  coherent (or fully correlated) and partly correlated source models with
\begin{equation}
\mathbf{O}=
\begin{bmatrix}
	 2 & 2 \\
	2 & 2
\end{bmatrix}~\text{and}~
\mathbf{O}=
\begin{bmatrix}
	 5 & 4 \\
	4 & 5
\end{bmatrix}.\nonumber
\end{equation}

\begin{algorithm}
\caption{Steepest Descent Based DOA Estimation} \label{alg1}
\begin{algorithmic}[1]
\STATE {$f(\boldsymbol{\beta})=-\mathcal{J}\big(\boldsymbol{\beta},\boldsymbol{\rho}^{(d)},\boldsymbol{\delta}^{(d)}\big)/L$, initialize $\boldsymbol{\beta}=\boldsymbol{\beta}^{(d-1)}\in\boldsymbol{\Gamma}$.}
        \WHILE{$\Vert\nabla f(\boldsymbol{\beta})\Vert_2>0.001$}
           \STATE {$t_v=\left\{
\begin{array}{ll}
-(\pi-\beta_v)/f'_v(\boldsymbol{\beta}),&f'_v(\boldsymbol{\beta})<0,\\
\beta_v/f'_v(\boldsymbol{\beta}),&f'_v(\boldsymbol{\beta})>0,\\
\infty,&f'_v(\boldsymbol{\beta})=0,
\end{array}
\right.
\forall v.$}
           \STATE {$t=0.1\times\min\{t_1,\dots,t_V\}.$}
           \WHILE{$f\big(\boldsymbol{\beta}-t\nabla f(\boldsymbol{\beta})\big)>f(\boldsymbol{\beta})-0.3t\Vert\nabla f(\boldsymbol{\beta})\Vert^2_2$}
           \STATE {$t=0.5t$.}
           \ENDWHILE
           \STATE {$\boldsymbol{\beta}=\boldsymbol{\beta}-t\nabla f(\boldsymbol{\beta})\in\boldsymbol{\Gamma}$.}
       \ENDWHILE
\STATE {$\boldsymbol{\beta}^{(d)}=\boldsymbol{\beta}$.}
\end{algorithmic}
\end{algorithm}




In Fig. \ref{fig1}, we compare the root mean square error (RMSE) performance of the ECME algorithm with the CRLB \cite{bibitem4}, \cite{bibitem5}. In addition, we also simulate the second space-alternating generalized EM (SAGE) algorithm for uncorrelated sources in \cite{bibitem20} and this SAGE algorithm adopts the same simulation settings in \cite{bibitem20}. Each RMSE is based on $2000$ independent trials and the two algorithms share the same initial point. We can see that as expected, the ECME algorithm obtains smaller RMSEs than the SAGE algorithm. More importantly, we can observe that the ECME algorithm attains the CRLB of $\boldsymbol{\beta}$ when the number of snapshots $L$ is large, which coincides with the well known conclusion that \emph{the stochastic CRLB of $\boldsymbol{\beta}$ can be achieved asymptotically by the stochastic ML estimator of $\boldsymbol{\beta}$} \cite{bibitem2}. Hence, the ECME algorithm is able to obtain the stochastic ML estimate of $\boldsymbol{\beta}$ in \eqref{eq3} given an accurate initial point.

In Fig. \ref{fig2}, we compare the ECME algorithm with two subspace based algorithms, which utilize the state-of-the-art subspace separation approaches in \cite{bibitem17}--\cite{bibitem18} and are called ``Approach 1+Root-MUSIC'' and ``Approach 2+Root-MUSIC'', respectively. The three algorithms process the same snapshots for each trial. We can observe that as expected, the ECME algorithm yields more closely spaced estimates of $(\beta_1,\beta_2)$ centered around $(50\degree,100\degree)$ since in DOA estimation, the ML technique offers the highest advantage in terms of accuracy.



\section{Conclusion}

In this letter, we employed and designed the ECME algorithm for stochastic ML direction finding, where the sources may be correlated, in unknown nonuniform noise. Theoretical analysis indicated that the ECME algorithm is computationally efficient and operationally stable. Simulation results confirmed the effectiveness of the algorithm.

\end{document}